\documentclass[twocolumn,aps,prd,amsmath,floats,floatfix,superscriptaddress,nofootinbib]{revtex4-1}
\usepackage[utf8]{inputenc}
\usepackage{amsmath}
\usepackage{amssymb}
\usepackage{graphicx}
\usepackage{appendix}
\usepackage[unicode=true,
 bookmarks=false,
 breaklinks=false,pdfborder={0 0 1},backref=false,colorlinks=false]
 {hyperref}

\makeatletter
\usepackage[normalem]{ulem}
\usepackage{verbatim}
\usepackage{mathrsfs}
\usepackage{mathtools}
\usepackage{amsfonts}
\usepackage{latexsym}
\usepackage{epsfig}
\usepackage{color}
\usepackage{graphicx,subfigure}
\usepackage{units}
\usepackage{slashbox}
\usepackage{envmath}
\usepackage{natbib}
\usepackage{bm}

\definecolor{orange}{rgb}{0.9,0.45,0}

\def\part_n{\partial_\perp}

\def\be{\begin{equation}}
\def\ee{\end{equation}}

\long\def\symbolfootnote[#1]#2{\begingroup%
\def\thefootnote{\fnsymbol{footnote}}\footnote[#1]{#2}\endgroup}

\makeatother

\begin{document}

\title{Black hole accretion of scalar clouds with spontaneous symmetry breaking}

\author{Sebastian Garcia-Saenz}
\affiliation{Department of Physics, Southern University of Science and Technology,
Shenzhen 518055, China}

\author{Guangzhou Guo}
\affiliation{Department of Physics, Southern University of Science and Technology,
Shenzhen 518055, China}

\author{Peng Wang}
\affiliation{Center for Theoretical Physics, College of Physics, Sichuan University,
Chengdu 610064, China}

\author{Xinmiao Wang}
\affiliation{Department of Physics, Southern University of Science and Technology,
Shenzhen 518055, China}
%\date{September 2024}

\begin{abstract}
Spontaneous scalarization of black holes typically occurs through the condensation of a scalar field, with the field evolving from a $U(1)$-symmetric phase into a symmetry-breaking one with lower energy. We show that there exist symmetry-breaking phases which are themselves unstable to the formation of an additional scalar condensate, or `cloud', which is partly accreted into the black hole. By studying the fully nonlinear dynamical evolution of the process, we find that symmetry breaking causes the accretion channels of scalar clouds to be non-degenerate, favoring a dominant channel for evolution. Additionally, the final states form a characteristic energy band due to varying amounts of radiation emitted by clouds in different channels.
\end{abstract}

\maketitle

\section{Introduction}

The past decade has witnessed spectacular progress in the field of black hole physics, driven primarily by the detection of gravitational waves from binary black hole mergers, which opened up new avenues for exploring the properties of black holes
\citep{Abbott:2016blz}. Prime among these is the testing of the validity of the black hole uniqueness theorems of general relativity (see \citep{Chrusciel:2012jk} for a review), particularly through the measurement of quasinormal
modes during the ringdown phase \citep{Price:2017cjr,Giesler:2019uxc}. The Event Horizon Telescope has similarly revolutionized our understanding of black holes by capturing the images of M87{*} and Sgr A{*} \citep{EventHorizonTelescope:2019dse,EventHorizonTelescope:2022wkp}, revealing a striking feature: a luminous ring encircling a dark shadow. These distinctive signatures have been attributed to the intense light deflection occurring near unstable bound photon orbits, providing insights into black hole physics in the strong field regime, e.g.\ the structure of luminous accretion disks \citep{Yuan:2014gma,Garcia:2013lxa,McKinney:2012vh,EventHorizonTelescope:2021srq,Blaes:2013toa}.

While these phenomena align with the predictions of general relativity \citep{LIGOScientific:2016lio,LIGOScientific:2017vox,LIGOScientific:2017bnn}, future observations of increased precision offer the exciting prospect of measuring deviations from Einstein's gravity interacting with standard matter \citep{LIGOScientific:2021sio}. For instance, the no-hair theorem \citep{Bekenstein:1971hc,Bekenstein:1995un} may be circumvented in the presence of a scalar field that couples non-minimally either to gravity or to other matter fields \citep{Herdeiro:2015waa,Sotiriou:2013qea,Antoniou:2017acq}. In this well-studied setup, black holes are subject to the growth of `hair' in the form of a scalar cloud, i.e.\ a nonstatic, but potentially long-lived, condensate of the field in the vicinity of the event horizon.\footnote{The term `scalar cloud', or more generally `boson cloud', appears to be mostly used in the context of rotating black holes, where a scalar condensate forms as a result of superradiance; see e.g.\ \citep{Zhang:2019eid,Baumann:2019eav,Baumann:2022pkl} for recent work and \citep{Sampaio:2014swa} for the electromagnetic analog. We apply this terminology here too, which is justified because the clouds formed via scalarization are also nonstatic and evolve through accretion and radiation. We emphasize however that, unlike in most previous studies, our numerical treatment of the cloud is fully nonlinear and does not neglect backreaction.} Although originally studied in the context of neutron stars \citep{Damour:1993hw}, this so-called spontaneous scalarization phenomenon has been shown to also arise in black hole spacetimes within scalar-tensor theories of gravity, the best-known model being scalar-tensor-Gauss-Bonnet theory and its extensions \citep{Doneva:2017bvd,Silva:2017uqg,Cunha:2019dwb,Herdeiro:2020wei,Berti:2020kgk}.

On the other hand, it has been appreciated that scalarization may occur in a simpler class of models, which feature no higher curvature terms and only minimal gravitational couplings, but they require the presence of an additional matter field with which the scalar interacts non-minimally. A very natural choice in this class is the Einstein-Maxwell-scalar (EMS) theory where the scalar field $\Phi$ couples to the standard Einstein-Maxwell Lagrangian through the term $f(|\Phi|)F^{\mu\nu}F_{\mu\nu}$, with $F_{\mu\nu}$ the electromagnetic field strength and $f$ some function \citep{Herdeiro:2018wub} (see also \citep{Stefanov:2007eq} for earlier work in the context of nonlinear electrodynamics). The spontaneous scalarization effect in the EMS model has been further studied in \citep{Fernandes:2019rez,Blazquez-Salcedo:2020nhs,Fernandes:2019kmh,Myung:2018vug}, while various properties of these scalarized black holes have been investigated in \citep{Gan:2021pwu,Gan:2021xdl,Guo:2021enm,Guo:2022ghl}, along with their spinning counterparts \cite{Guo:2023ivz,Guo:2024bkw}.

Scalarization in the EMS model occurs dynamically through a tachyonic (i.e.\ negative effective potential) instability of an initial scalar field perturbation on the background of a hairless black hole, driving the system into a new configuration with lower energy accompanied by a non-trivial scalar field distribution. In this paper we ignore spin, so the initial state is a Reissner-Nordstr\"om (RN) black hole; notice that electric charge is needed to trigger the destabilization effect \citep{Herdeiro:2018wub}. The appearance of a nontrivial scalar profile, or condensate, may be naturally understood as a process of spontaneous symmetry breaking. In the original setup, the scalar field is real, so the symmetry that is broken by the scalarization effect is $\mathbb{Z}_2$, or $\Phi\mapsto-\Phi$. Here we focus instead on the case of a complex scalar as it provides a more interesting symmetry breaking pattern, namely that of a (global) $U(1)$ symmetry. The study of spontaneous scalarization in this scenario has been pioneered in \citep{Latosh:2023cxm,Hyun:2024sfv}.

In this paper, we employ numerical relativity to investigate the fully nonlinear dynamics of the EMS model, specifically the spontaneous scalarization of a charged black hole and the subsequent evolution of the scalar cloud. After its initial formation following the tachyonic destabilization, we observe that the cloud continues to evolve, on a timescale of order the distance between the cloud and the event horizon, in a process of energy loss both to accretion into the black hole and ejection to spatial infinity. We find essential differences between the scenario where scalarization occurs starting in a symmetric phase, i.e.\ without further instabilities upon formation of a scalar condensate, versus the case where a scalar cloud forms from a symmetry-broken phase. Unlike with scalarization from the symmetric phase, the breaking of $U(1)$ symmetry allows for multiple accretion channels for the cloud, each leading to different amounts of energy loss and distinct final states. Notably, a dominant accretion channel emerges due to symmetry breaking.

\section{Scalarization}

We consider the EMS model of \citep{Herdeiro:2018wub}, extended here to a complex scalar field $\Phi=\phi_{1}+i\phi_{2}$, which is non-minimally coupled to the electromagnetic field $A_{\mu}$. The complete action is (we use geometrized units with $G=c=1$)
\begin{equation}
S=\frac{1}{16\pi}\int d^{4}x\sqrt{-g}\left[R-2\partial_{\mu}\Phi^{*}\partial^{\mu}\Phi-f\left(\left|\Phi\right|\right)F^{\mu\nu}F_{\mu\nu}\right], \label{eq:action}
\end{equation}
where $F_{\mu\nu}=\partial_{\mu}A_{\nu}-\partial_{\nu}A_{\mu}$. In this paper, we adopt $f\left(\left|\Phi\right|\right)=e^{\alpha|\Phi|^2}$ for the non-minimal coupling function, with $\alpha$ a real constant, a choice which admits an electrovacuum solution given by a RN black hole and $\Phi=0$.\footnote{More generally, the function $f(|\Phi|)$ must satisfy $f''>0$ and $|\Phi|f'>0$ (here primes denotes derivatives of $f$ with respect to its argument) for some range of the radial coordinate~\citep{Herdeiro:2018wub}. In addition, $f'(0)=0$ in order for the model to admit hairless solutions.} Note that we do not include a minimal coupling of the scalar field to electromagnetism, i.e.\ $\Phi$ is uncharged under the electromagnetic $U(1)$, unrelated to the global $U(1)$ symmetry $\Phi\mapsto e^{i\varphi}\Phi$. The electrovacuum solution preserves the $U(1)$ symmetry, so we refer to it as the symmetric phase.

A small scalar perturbation $\delta\Phi$ on the RN black hole background obeys the equation $\left(\Box-\mu_{\rm eff}^{2}\right)\delta\Phi=0$, where $\mu_{\rm eff}^{2}\equiv -\alpha Q^{2}/r^4$ ($r$ is the radius in standard spherical coordinates) and $Q$ is the charge of the hole. If $\alpha>0$, the negative value of $\mu_{\rm eff}^{2}$ introduces a tachyonic effect, potentially destabilizing the background and triggering scalarization.
\begin{figure}[t!]
\centering{}\includegraphics[width=1\linewidth]{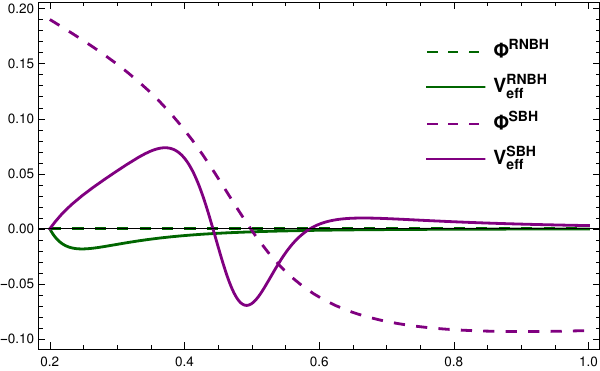}\caption{Scalar field profiles (dashed lines) and effective potentials (solid) for $Q/M=0.55$ RN (green) and $Q/M=1.29$ scalarized (purple) black holes, respectively. The event horizons are set at $r_{h}=0.2$. Both the RN and scalarized black holes (SBH) possess negative potential wells. In the latter case, the well is separated from the horizon by a potential barrier, and its location coincides with the node of the scalar field.}
\label{Veff}
\end{figure}

More in detail, and considering now a general static and spherically symmetric ansatz for the metric, $ds^{2}=-e^{-2\delta(r)}N(r)dt^{2}+dr^{2}/N(r)+r^{2}d\Omega^{2}$, one finds the equation $\frac{d^{2}\Psi}{dx^{2}}+\left(\omega^{2}-V_{\rm eff}(x)\right)\Psi=0$ obeyed by the spherical, or monopole, perturbation $\Psi\equiv r\delta\Phi$, where $\omega$ is the frequency, $x$ is defined by $dx/dr\equiv e^{\delta(r)}/N(r)$ and $V_{\rm eff}$ is the effective potential.\footnote{Explicitly,
\begin{equation*}
V_{\rm eff}=\frac{N}{r^{2}e^{2\delta}}\left[1-N-2r^{2}\left|\Phi^{\prime}\right|^{2}-\frac{Q^{2}\left(1+\alpha-2\left|\alpha\Phi-r\Phi^{\prime}\right|^{2}\right)}{r^{2}e^{\alpha\left|\Phi\right|^{2}}}\right].
\end{equation*}
} In Fig.\ \ref{Veff}, the effective potential for the $U(1)$-symmetric phase is seen to display a negative well near the event horizon, and a numerical calculation shows that a tachyon mode with $\omega=0.2454iM$ is present. In agreement with expectations, this mode is responsible for triggering spontaneous scalarization \citep{Herdeiro:2018wub,Guo:2024cts}. Intriguingly, we find that there exist $U(1)$-broken phases (scalarized black holes with $\Phi\neq0$), which also exhibit a negative potential well, a feature that is linked to the presence of a node in the scalar profile. This suggests that these symmetry-breaking phases are unstable to the formation of a scalar cloud localized around the potential well. Our aim in this paper is to advance the investigation of the nonlinear dynamics of scalar clouds to encompass both $U(1)$-symmetric and $U(1)$-broken phases.

\section{Dynamical formation of scalar clouds}

To simulate the fully nonlinear evolution in the EMS model, we use the Baumgarte-Shapiro-Shibata-Nakamura (BSSN) formulation implemented in \textit{BlackHoles@Home} \citep{Bhathome}; see \citep{Bona:2003fj,Brown:2009dd,Etienne:2014tia,Ruchlin:2017com,Etienne:2024ncu,Baumgarte:2012xy} for references and Appendix \ref{app:A} for some technical details and numerical convergence tests. To ensure consistent numerical results, we also perform nonlinear simulations with the \textit{Einstein Toolkit} \citep{Zachariah}. Given our focus on spherically symmetric evolution, we have the electromagnetic field $A_{\mu}=A(t,r)\delta^t_{\mu}$ (choosing the gauge $A_r=0$), and from the equations of motion one infers $\partial_{\mu}\left(\sqrt{-g}e^{\alpha\Phi^{*}\Phi}\partial^{r}A^{t}\right)=0$, indicating that the black hole charge $Q$, defined by the integration constant, is conserved during temporal evolution. For our numerical study, we set units such that $Q=1$ and $\alpha=160$.\footnote{Although the tachyonic instability does not require very large values of $\alpha$, such a choice makes the instability rate faster and thus easier to simulate numerically.} For the initial data, we introduce a scalar perturbation $\delta\Phi=pe^{-\frac{\left(r-r_{0}\right)^{2}}{\Delta^{2}}}e^{i\theta_{0}}$ around both the $U(1)$-symmetric and $U(1)$-broken phases. Here, $\theta_{0}$ is the phase factor of the perturbation, the amplitude is chosen as $p=10^{-6}$, the location $r_{0}$ is set around the negative potential well, and the width is chosen as $\Delta=1$. For latter use, we define $\Phi_{h}$ to be the (time-dependent) value of the scalar field $\Phi$ at the event horizon. At the background level, i.e.\ prior to the introduction of the above perturbation $\delta\Phi$, the $U(1)$-breaking system is chosen, without loss of generality, to have a real and positive value of $\Phi$.
\begin{figure}[t!]
\centering{}\includegraphics[height=0.45\textheight]{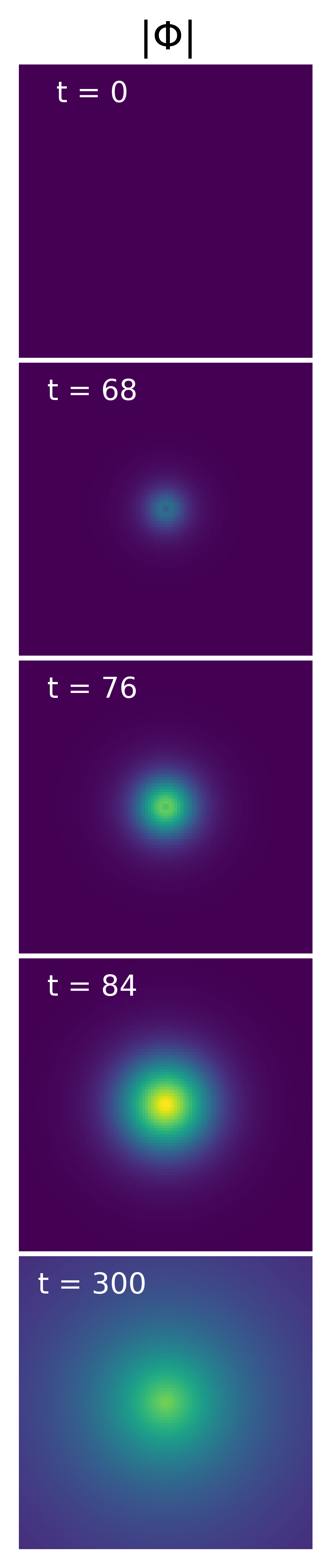}\includegraphics[height=0.45\textheight]{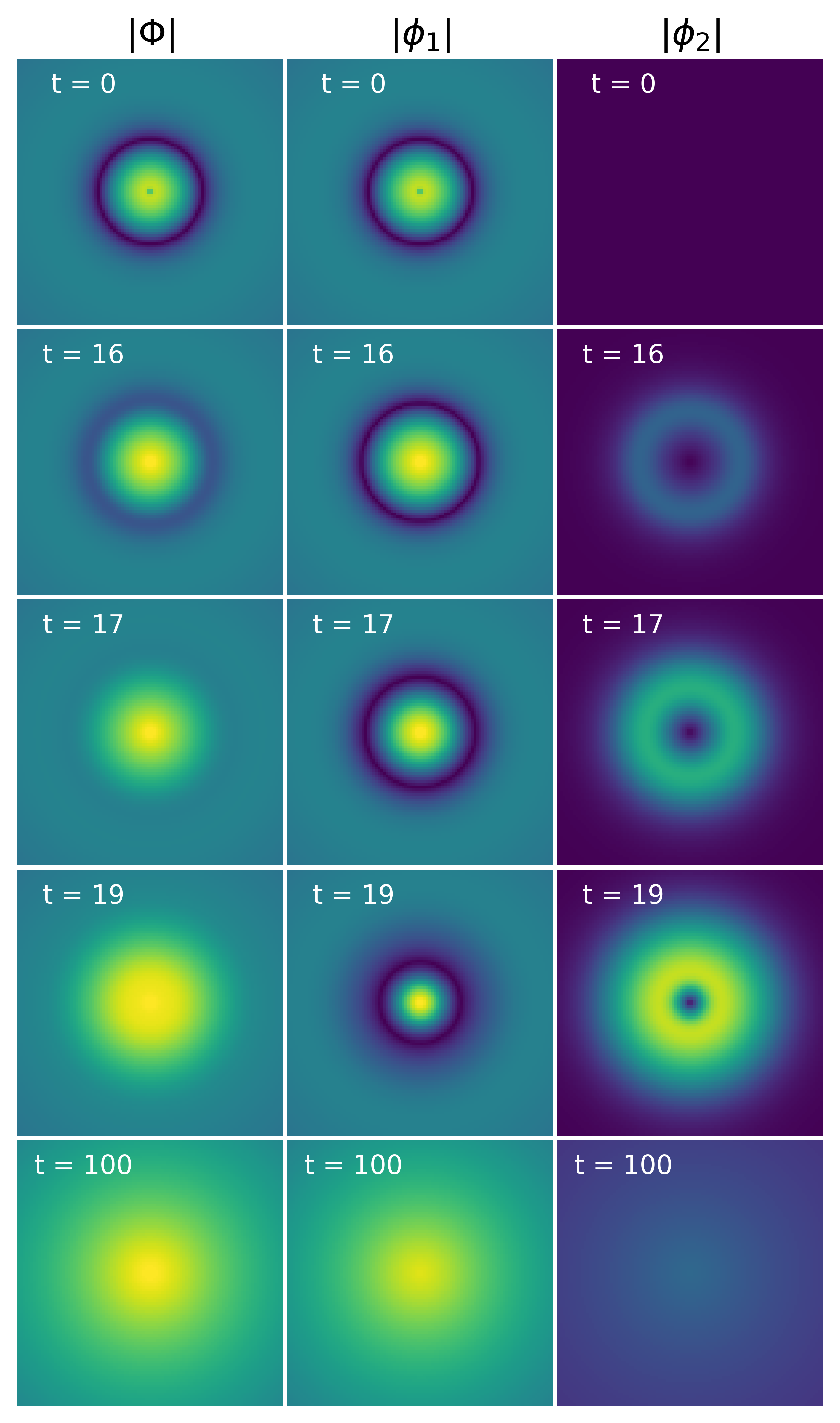}
\includegraphics[height=0.048\textheight]{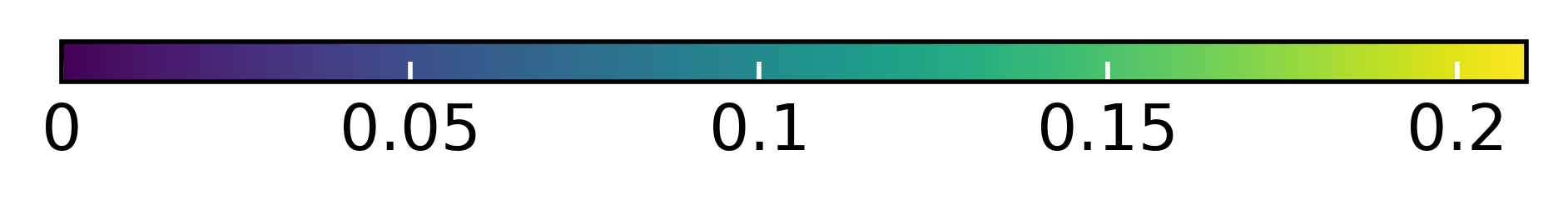}\caption{Dynamical formation of a scalar cloud for the initial perturbation with $\theta_{0}=\pi/2$ in the $U(1)$-symmetric ($M=2.5$, leftmost column) and $U(1)$-broken ($\Phi_{h}=0.19$ and $M=0.7744$, right columns) phases. A scalar cloud develops within the negative potential well due to scalarization and is subsequently accreted onto the black hole, terminating in the formation of a stable cloud structure around it.
}
\label{SCtheta90}
\end{figure}

In Fig.\ \ref{SCtheta90}, an initial wavepacket with phase $\theta_{0}=\pi/2$, perturbing only the field component $\phi_{2}$, is placed within the negative potential well. In the symmetric phase, the field amplitude $\left|\Phi\right|$ grows near the event horizon, forming a scalar cloud within the potential well. Ultimately, a long-lived cloud accretes onto the black hole, with the evolution terminating in a scalarized black hole, consistent with the results of \citep{Herdeiro:2018wub,Guo:2024cts}. In the $U(1)$-broken phase, the negative potential well is separated from the horizon of the scalarized black hole (with $\Phi_{h}=0.19$ in our simulation). The perturbation grows via tachyonic instabilities, forming a ring structure for $\phi_{2}$ around the well (see the rightmost column). After approximately $t\approx 20$ in our numerics, the cloud becomes massive enough and is drawn into the black hole by the gravitational pull. As a result, the cloud component $\phi_{2}$ is gradually accreted onto the hole, forming a lighter cloud that remains near the horizon in equilibrium, i.e.\ a hairy black hole. During this process, the node of the scalar field $\left|\Phi\right|$ gradually fades as the $\phi_{2}$ cloud forms and eventually disappears.
\begin{figure}[t!]
\centering{}\includegraphics[width=0.9\linewidth]{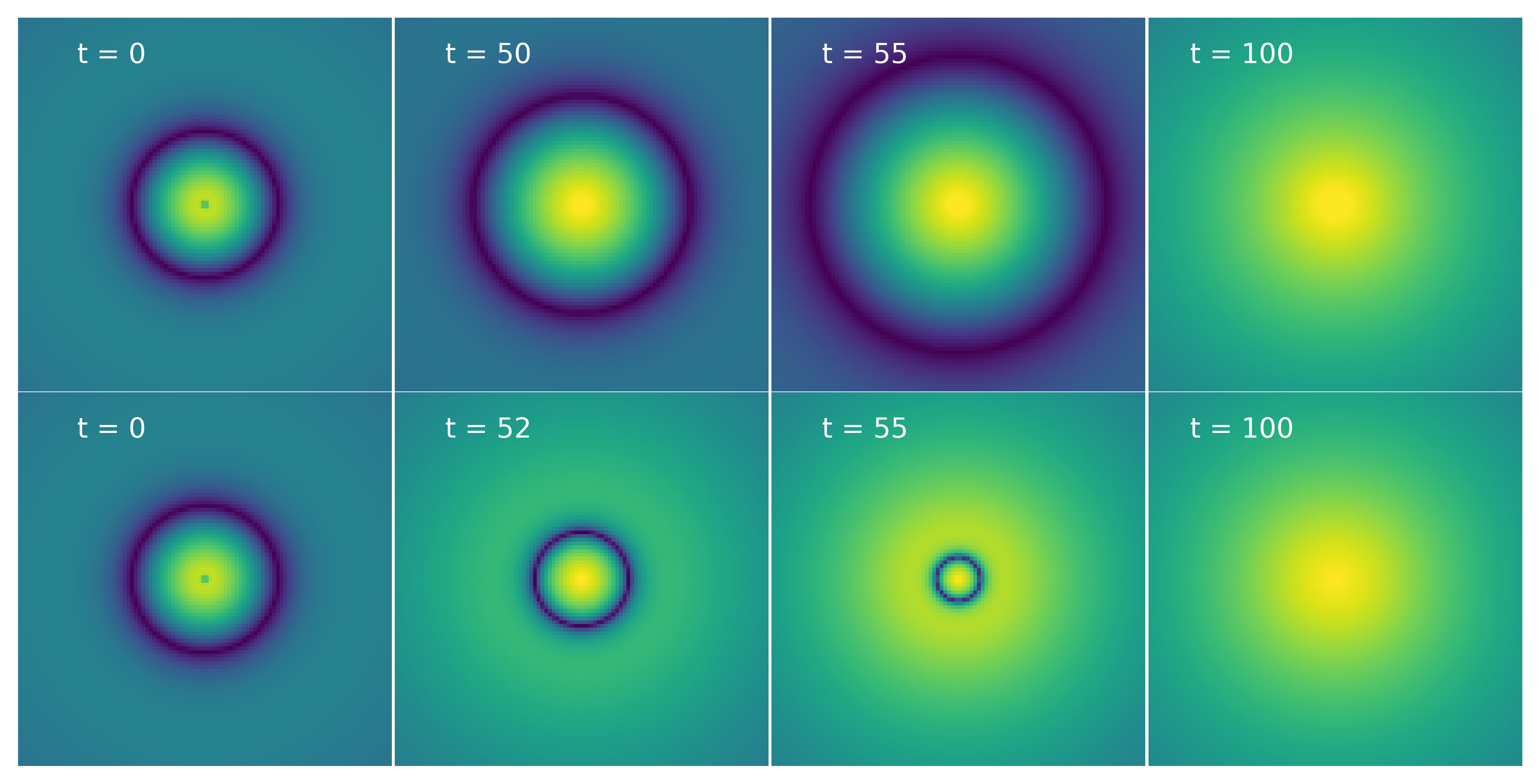}\includegraphics[width=0.1\linewidth]{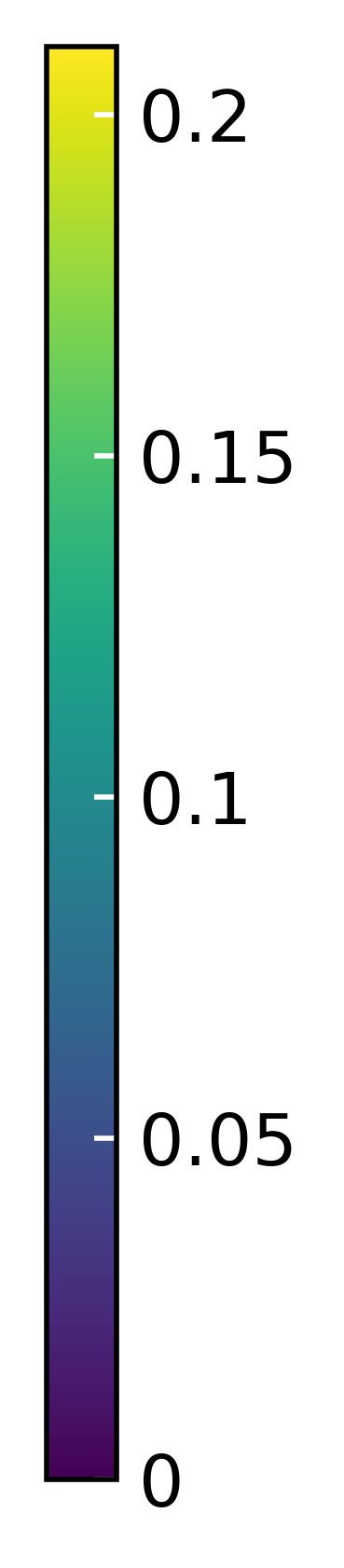}\caption{Dynamical evolution of the field norm $\left|\Phi\right|$ for initial perturbations with $\theta_{0}=0$ (upper row) and $\theta_{0}=\pi$ (lower) in the symmetry-broken phase ($\Phi_{h}=0.19$ and $M=0.7744$). The scalar cloud formed from the positive perturbation ($\theta_{0}=0$) can induce nonlinear effects at late times, causing the scalar field node to propagate towards spatial infinity. Conversely, the negative perturbation ($\theta_{0}=\pi$) leads to the node of the field moving into the event horizon at late times, resulting in a stable, nodeless scalar distribution outside the black hole.}
\label{theta0pi}
\end{figure}
\begin{figure}[t!]
\centering{}\includegraphics[width=1\linewidth]{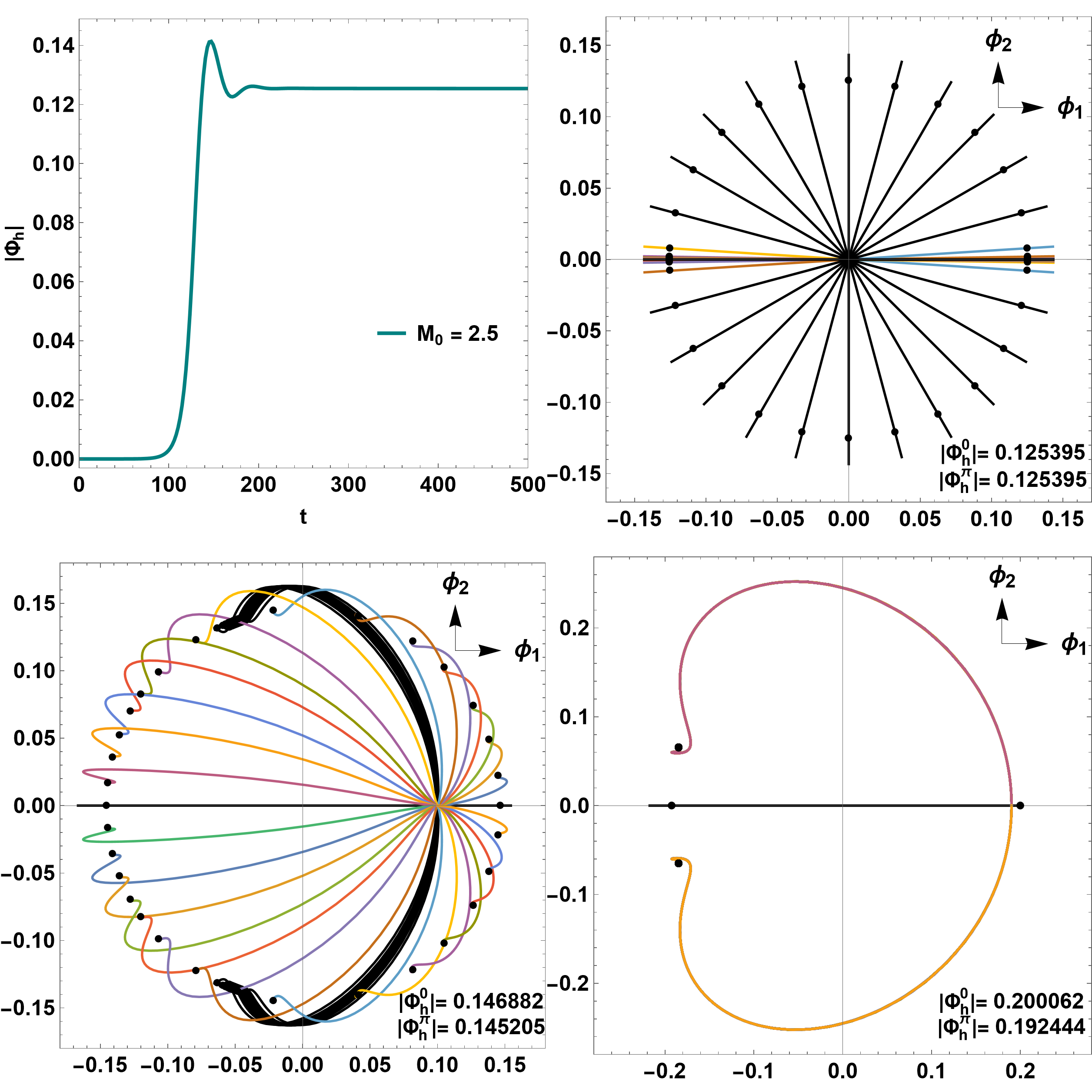}\caption{Channels of scalar cloud accretion, represented by the value of the scalar field at the event horizon, $\Phi_{h}$. \textit{Upper row:} In the $U(1)$-symmetric phase ($\Phi=0$ and $M=2.5$), the accretion dynamics and final states remain $U(1)$-symmetric with respect to the initial perturbation phase $\theta_{0}$. \textit{Lower:} In the $U(1)$-broken phases, using here $\Phi_{h}=0.1$, $M=1.8744$ (left panel) and $\Phi_{h}=0.19$, $M=0.7744$ (right panel), the accretion paths no longer respect the $U(1)$ symmetry. Cloud accretion tends to follow a dominant path, a phenomenon that becomes more pronounced for larger $\Phi_{h}$ values of the initial field: in the right panel, the black and colored lines almost completely overlap (the visible purple and orange lines cover all other colored as well as black lines). As a result, the final states are not $U(1)$-symmetric, as indicated by the varying values of $\left|\Phi_{h}\right|$ shown in the lower right corners.
}
\label{PHih}
\end{figure}

While in the symmetric phase the dynamics remain $U(1)$-equivalent, in the symmetry-broken phase the accretion process can vary significantly based on the initial perturbation phase. In Fig.\ \ref{theta0pi}, we exhibit the evolution of the scalar field for perturbations with $\theta_{0}=0$ and $\theta_{0}=\pi$, neither of which induces a scalar cloud in the $\phi_{2}$ direction. For the $\theta_{0}=0$ perturbation, a cloud of positive $\phi_{1}$ accumulates within the potential well. As the cloud grows, non-linear effects cause the node of $\Phi$ to propagate towards spatial infinity. A similar scenario occurs for the $\theta_{0}=\pi$ perturbation, now with a cloud of negative $\phi_{1}$, driving the node of $\Phi$ inward towards the black hole. In both cases, the formation and accretion of scalar clouds effectively peel off the negative potential well and ``radiate'' the node of the field.

\begin{figure}[t!]
\begin{centering}
\includegraphics[width=1\linewidth]{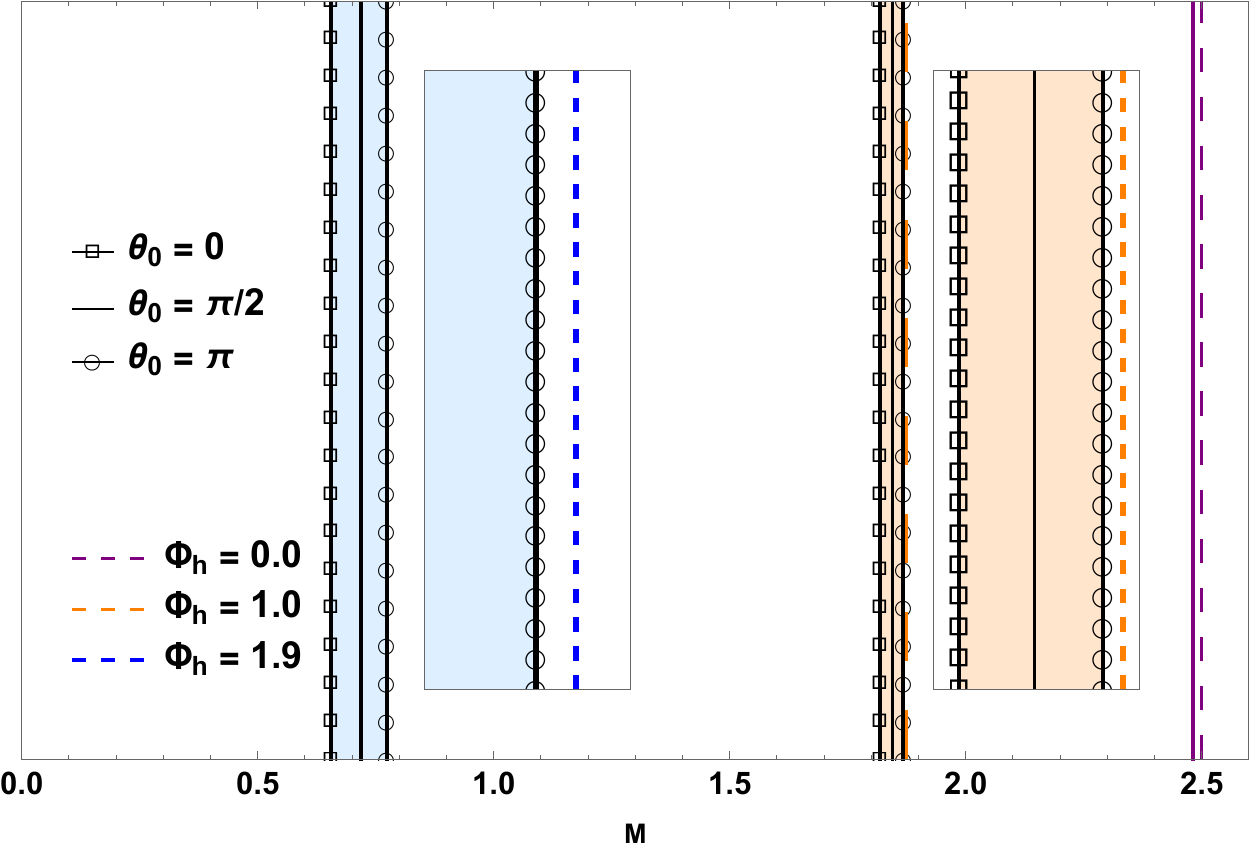}
\par\end{centering}
$\,$
\begin{centering}
\includegraphics[width=1\linewidth]{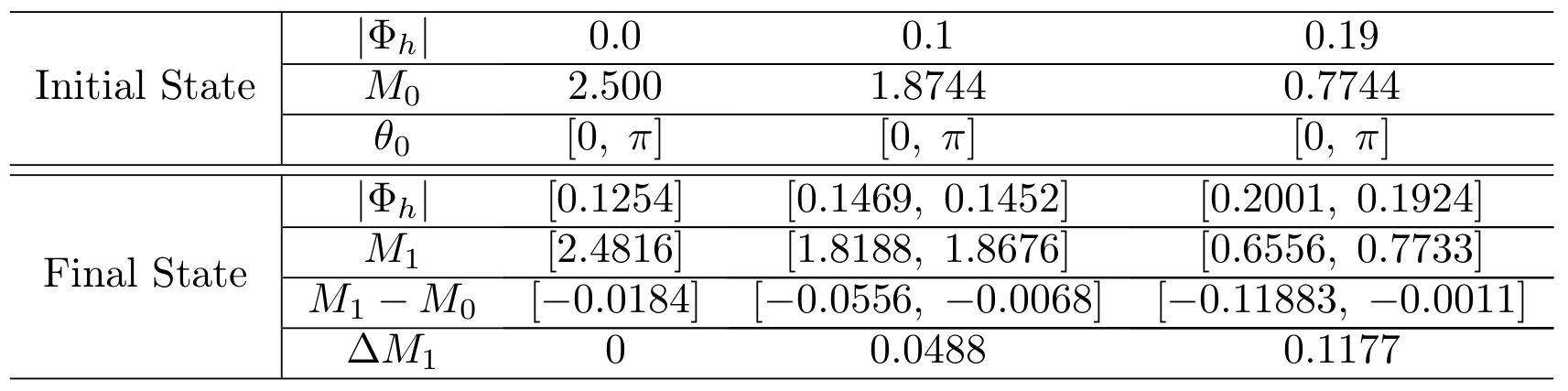}
\par\end{centering}
\centering{}\caption{\textit{Upper panel:} Energy of initial (dashed lines) and final (solid) states. Energy loss in the $U(1)$-symmetric phase is illustrated by the purple lines, and we consider two $U(1)$-breaking models (orange and blue). In the latter cases, the colored bands represent the range of final values of the energy. The bands are bounded by the cases with $\theta_0=0$ (squares) and $\theta_0=\pi$ (circles).
%The colored and black solid lines match those represented in Fig.\ \ref{PHih}.
The zoomed-in insets allow one to discern the energy gap between initial and final states. \textit{Lower:} The table provides explicit data for both the initial and end states in our numerics, in particular the energy loss. Here, $M_{0}$ represents the energy of the initial state, $M_{1}$ is the energy of the end state, and $\Delta M_{1}$ denotes the width of the energy band for the final states.
}
\label{Energy}
\end{figure}

To further explore the relationship between the dynamical accretion of scalar clouds and the initial perturbation phase $\theta_{0}$, we systematically examine the evolution of the scalar field at the event horizon, $\Phi_{h}(t)$, for a range of $\theta_{0}$ values. In Figs.\ \ref{PHih} and \ref{Energy}, 24 uniformly spaced values (black lines) within the interval $(0,2\pi)$, as well as a set of specific values (colored lines) within the narrower intervals $(-\pi/12,\pi/12)$ and $(11\pi/12,13\pi/12)$. In the symmetric phase, we observe that the accretion process of the cloud manifests the $U(1)$ symmetry, as depicted in the upper row of Fig.\ \ref{PHih}, which shows in particular the conservation of the phase $\theta$ during time evolution. On the other hand, in the $U(1)$-broken phase, the symmetry breaking causes the phase factor $\theta$ to become a non-trivial dynamical variable during the accretion, as illustrated in the lower panels of Fig.\ \ref{PHih}. We observe a manifest loss of $U(1)$ invariance in the process, although it remains symmetric with respect to the $\phi_{1}$-axis due to a residual $\mathbb{Z}_2$ symmetry.\footnote{The $\phi_1$-axis is preferred because of our choice of the background $\Phi$ as having a vanishing $\phi_2$ component. More generally, the axis of symmetry will be defined by the direction of $\Phi$ in the field space.} Interestingly, the symmetry breaking introduces a preferred path for scalar cloud accretion, acting as an attractor that draws in the majority of accretion channels. Fig.\ \ref{SCtheta90} illustrates one such typical path ($\theta_0=\pi/2$). Furthermore, the final equilibrium state will not be $U(1)$-symmetric for scalar clouds in the symmetry-broken phase. As shown in the lower right corner of the lower panels, we find that $\left|\Phi_{h}^{\theta_0}\right|$, which represents the value of $\left|\Phi_{h}\right|$ at the end state for a given initial perturbation phase $\theta_{0}$, is maximal for $\theta_0=0$ and minimal for $\theta_{0}=\pi$ (see also Fig.\ \ref{Energy}).

The existence of a dominant path in field space may be understood from the fact that one expects most of the energy of the field to go into the gapless degrees of freedom, in this case the angular component of $\Phi$, with a comparatively smaller fraction in the radial field.\footnote{We would like to thank an anonymous referee for raising this interesting question.} This picture is consistent with the observed trajectory, which appears to be roughly circular, at least qualitatively. This intuition is corroborated by an explicit comparison of the energy densities stored in the radial and angular components of the field; see Appendix \ref{app:B} for details.

\section{Energy loss}

A noteworthy phenomenon during scalar cloud accretion is that the cloud reaches equilibrium after a portion of its energy is radiated to spatial infinity (see Figs.\ \ref{SCtheta90} and \ref{theta0pi}), potentially causing the final state to have less energy than the initial state. In Fig.\ \ref{Energy}, we calculate the Arnowitt-Deser-Misner mass \citep{Misner:1973prb} of the initial and final states to quantify the energy loss during accretion in both the $U(1)$-symmetric and $U(1)$-breaking phases. In both cases, a discernible energy gap exists between the initial and final states, confirming that the formation of a stable scalar cloud involves the radiation
of energy to spatial infinity.

In the symmetric phase, the end states maintain the $U(1)$-symmetry and are degenerate with with respect to the final energy ($M_{1}=2.4816$ in our numerics). The energy loss, calculated as $M_{1}-M_{0}$ (displayed in the table of Fig.\ \ref{Energy}), indicates a transition from the $U(1)$-symmetric phase to a lower-energy final state. On the other hand, in the $U(1)$-broken phase, symmetry breaking lifts this degeneracy and leads to an energy band for the final states, bounded by the cases with initial phases $\theta_{0}=0$ and $\theta_{0}=\pi$. Notably, the $\theta_{0}=0$ perturbation induces the formation of a positive $\phi_{1}$ cloud, with most of its energy radiating to spatial infinity. It raises a backreaction on the field itself, which pushes the node of $\Phi$ outward from the potential well towards spatial infinity, as mentioned previously. As a result, the end state for $\theta_{0}=0$ has the minimal energy. Conversely, for the $\theta_{0}=\pi$ perturbation, most of the negative $\phi_{1}$ cloud propagates toward the black hole, driving the node of $\Phi$ into the hole to reach equilibrium. In this scenario, a minimal amount of energy from the cloud radiates to infinity, leading to the end state having maximal energy. For detailed numerical data, refer to the table in Fig.\ \ref{Energy}.

\section{Conclusions}

In the EMS model, the non-minimal coupling between the complex scalar and electromagnetic fields can induce a negative potential well near a RN black hole ($U(1)$-symmetric phase), triggering the onset of scalarization and leading to the accretion of a scalar cloud onto the event horizon. During dynamical evolution, the accretion of scalar clouds preserves the global $U(1)$ symmetry, resulting in $U(1)$-symmetric final states with energy that is degenerate with respect to the initial perturbation phase $\theta_{0}$. Remarkably, even in the $U(1)$-breaking phase, a negative potential well may exist in the presence of a non-trivial scalar field. Perturbations with different initial phases can induce the formation of scalar clouds within this well, which subsequently accrete through the event horizon, bringing energy to the black hole and radiating some energy to spatial infinity. Unlike the symmetry-preserving case, symmetry breaking allows for multiple accretion channels for scalar clouds with different initial phases. This phenomenon leads to varying amounts of energy radiated to spatial infinity, resulting in a lifting of degeneracy and the appearance of a band of possible final states. Moreover, symmetry breaking establishes a dominant accretion channel, acting as an attractor which draws in the majority of field trajectories.

This paper highlights the significant impact of spontaneous symmetry breaking on the dynamics of scalar cloud accretion. Our studies reveal distinctive signatures of symmetry breaking in black holes, providing compelling motivation for their exploration in astrophysical observations. The assumption of spherical symmetry in our setup is admittedly simplistic, and implies that neither electromagnetic nor gravitational radiation will be emitted during the evolution. It would be therefore interesting to relax this assumption in order to investigate these radiation channels and how the properties of the system may be assessed through their observation. Future research could also extend these findings to more realistic systems, such as rotating black holes, where the superradiance phenomenon is likely to produce richer scalar cloud structures and dynamics.

\bigskip{}

\noindent \textbf{Acknowledgements.} We are grateful to Yupeng Zhang and Shenkai Qiao for useful discussions and valuable comments. SGS, GG and XW are supported by the NSFC (Grant Nos.\ 12250410250 and 12347133). PW is supported in part by the NSFC (Grant Nos.\ 12105191, 12275183, 12275184 and 11875196).

\appendix

\section{Numerical scheme for black hole evolution}
\label{app:A}

In this paper, we simulate the black hole evolution using the $3+1$ decomposition of the metric, expressed as
\begin{equation}
	ds^{2}=-N_{0}^{2}dt^{2}+\gamma_{ij}\left(dx^{i}+N^{i}dt\right)\left(dx^{j}+N^{j}dt\right) \,.
\end{equation}
For the gravity sector, we apply the BSSN formulation, using the $\textrm{1+log}$ slicing and Gamma-driver shift conditions \citep{Bona:2003fj,Brown:2009dd,Etienne:2014tia,Ruchlin:2017com,Etienne:2024ncu,Baumgarte:2012xy}. 

To model the scalar field, we employ dynamical variables $\Phi$ and $\Pi$, where $\Pi=n^{\mu}\nabla_{\mu}\Phi$, where $n_{\mu}$ is given by $n_{\mu}=\left(-N_{0},0,0,0\right)$, as the 4-vector orthogonal to the spatial hypersurface. For the electromagnetic field, we decompose $F_{\mu\nu}$ into the 3-dimensional electric and magnetic components, defined as $E_{i}=\gamma_{i}^{\mu}n^{\nu}F_{\mu\nu}$ and $B_{i}=\gamma_{i}^{\mu}n^{\nu}*F_{\mu\nu}$,
respectively, where $*$ denotes the Hodge dual, and $\gamma_{i}^{\mu}$ projects 4-dimensional vectors onto the spatial hypersurface. Since we focus on spherically symmetric evolution, the magnetic field
$B_{i}$ vanishes during the simulation. The evolution equations for the matter fields read
\begin{align}
	\partial_{\perp}\Phi & =N_{0}\Pi \,,\nonumber \\
	\partial_{\perp}\Pi & =D_{i}\left(N_{0}\chi^{i}\right)+N_{0}\Pi K-\frac{1}{2}\alpha N_{0}\Phi e^{\alpha|\Phi|^{2}}F^{2} \,,\nonumber \\
	\partial_{\perp}E^{i} & =\alpha KE^{i}-2\alpha N_{0}E^{i}\left(\phi_{1}\Pi_{1}+\phi_{2}\Pi_{2}\right) \,,\label{eq:eq of matter}
\end{align}
where the time derivative is defined by $\partial_{\perp}\equiv\partial_{t}-\mathcal{L}_{N}$,
and $\mathcal{L}_{N}$ is the Lie derivative along $N_{i}$; $K$ is the trace of the extrinsic curvature, $F^2=-2E^iE_i$, and $\phi_{1,2}$ and $\Pi_{1,2}$ denote respectively the real and imaginary parts of $\phi$ and $\Pi$.
\begin{figure}[t!]
\centering{}\includegraphics[width=1\linewidth]{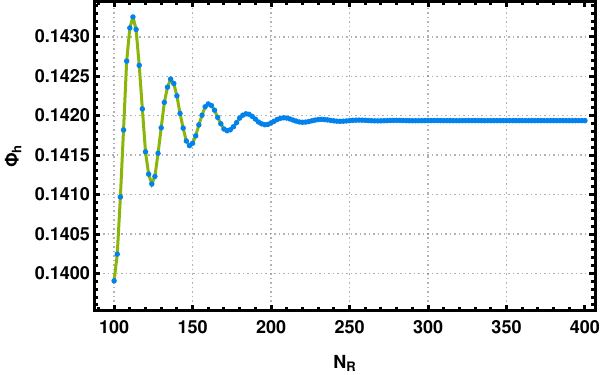}
\caption{Convergence test of the numerical scheme, illustrated by the event-horizon value $|\Phi_h|$ of the scalar field as function of the number $N_R$ of grid points. The blue dots represent the final state of spontaneous scalarization from a RN black hole with $Q/M=0.5$ and $\alpha=160$. As the grid number increases, $|\Phi_h|$ converges, indicating convergence of the simulation.}
\label{conTest}
\end{figure}

During the evolution, the gravity-matter interaction is described by the stress-energy tensor $T_{\mu\nu}$, given by
\begin{align}
	T_{\mu\nu}&= \frac{1}{8\pi}\left(2\partial_{(\mu}\Phi^{*}\partial_{\nu)}\Phi-g_{\mu\nu}\left|\partial\Phi\right|^{2}\right)\nonumber \\
	&\quad +\frac{1}{8\pi}e^{\alpha|\Phi|^{2}}\left(2F_{\mu\rho}F_{\nu}{}^{\rho}-\frac{1}{2}g_{\mu\nu}F^{2}\right) \,.
\end{align}
To incorporate the matter contribution into the BSSN formalism, we project the stress-energy tensor onto $3+1$ variables, 
\begin{align}
	\rho&= n^{\mu}n^{\nu}T_{\mu\nu}=\frac{1}{8\pi}\gamma^{ij}\left(\partial_{i}\phi_{1}\partial_{j}\phi_{1}+\partial_{i}\phi_{2}\partial_{j}\phi_{2}\right) \nonumber \\
	&\quad +\frac{1}{8\pi}\left(\Pi_{1}\Pi_{1}+\Pi_{2}\Pi_{2}e^{\alpha|\Phi|^{2}}E^{2}\right) \,, \nonumber \\
	J_{i}&= -\gamma_{i}^{\mu}n^{\nu}T_{\mu\nu}=-\frac{1}{4\pi}\left(\partial_{i}\phi_{1}\Pi_{1}+\partial_{i}\phi_{2}\Pi_{2}\right) \,, \nonumber \\
S_{ij}&= \gamma_{i}^{\mu}\gamma_{j}^{\nu}T_{\mu\nu} \nonumber\\
 &=\frac{1}{4\pi}\left(\partial_{i}\phi_{1}\partial_{j}\phi_{1}+\partial_{i}\phi_{2}\partial_{j}\phi_{2}-e^{\alpha|\Phi|^{2}}E_{i}E_{j}\right) \nonumber \\
	&\quad -\frac{1}{8\pi}\gamma_{ij}\gamma^{kl}\left(\partial_{k}\phi_{1}\partial_{l}\phi_{1}+\partial_{k}\phi_{2}\partial_{l}\phi_{2}\right) \nonumber \\
	&\quad +\frac{1}{8\pi}\gamma_{ij}\left(\Pi_{1}\Pi_{1}+\Pi_{2}\Pi_{2}+e^{\alpha|\Phi|^2}E^{2}\right) \,.
 \label{eq:stress energy}
\end{align}

In our numerical simulations, we adopt a spherical-like coordinate system, introducing a dimensionless radial coordinate $R$ to replace the standard spherical radial coordinate $r$. The relationship between $r$ and $R$ is defined as
\begin{equation}
	r=r_{\rm max}\left(R_{0}+\frac{e^{R/a}-e^{-R/a}}{e^{1/a}-e^{-1/a}}\right) \,,
\end{equation}
where $r_{\rm max}$ denotes the outer boundary, and $R_{0}$ and $a$ are scaling constants that map between $r$ and $R$. Since we focus on a spherically symmetric spacetime, the radial direction $R$ is discretized uniformly with $N_{R}$ grid points, while $N_{\theta}=N_{\varphi}=1$ for the angular directions $\left(\theta,\varphi\right)$. 

In Fig.\ \ref{conTest}, we present a convergence test for the simulation of spontaneous scalarization of a RN black hole, evaluated across different grid resolutions $N_{R}$. In the simulation setup, the radial coordinate parameters are specified
as $r_{\rm max}=30000$, $r_{0}=0.00012$ and $a=0.07$ (recall that we use units such that $Q=1$). The integration time step $\Delta t$ is chosen to match the smallest spatial $\Delta r_{\rm min}$ (i.e., $\Delta t=\Delta r_{\rm min}$) to avoid numerical instabilities. Our numerical results indicate that the simulation converges with increasing grid number $N_{R}$.

\section{Radial-angular scalar field decomposition}
\label{app:B}

In this Appendix we expand on the relationship between the dominant path observed in the scalar cloud accretion channels and the question of energy distribution among the field components. At a qualitative level, and at first approximation, the attractor trajectory in Fig.\ \ref{PHih} (black curves in bottom left panel) appears to correspond to circular motion, suggesting that the energy density of the scalar field is predominantly in the angular field. This agrees with the expectation that most of the energy should go to the gapless degrees of freedom in the system, with a comparatively smaller amount going to the radial part of the scalar field.

We can confirm this intuition through an explicit calculation of the energy density of each field component, with the radial and angular fields defined via
\begin{equation}
\Phi=\sigma(x) e^{i \theta(x)} \,.
\end{equation}
We define the stress-energy tensors associated to $\sigma$ and $\theta$ by
\begin{align}
    T^{(\sigma)}_{\mu\nu}&\equiv T_{\mu\nu}\Big|_{\theta=\theta_0,A=0} \nonumber\\
    &=\frac{1}{8\pi}(2\partial_\mu\sigma\partial_\nu\sigma-g_{\mu\nu}\partial^\rho\sigma\partial_\rho\sigma)\,, \nonumber\\
    T^{(\theta)}_{\mu\nu}&\equiv T_{\mu\nu}\Big|_{\sigma=\sigma_0,A=0} \nonumber \\
    &=\frac{1}{8\pi}(2\sigma_0^2 \theta^2\partial_\mu\theta\partial_\nu\theta-g_{\mu\nu}\theta^2\sigma_0^2\partial^\rho\theta\partial_\rho \theta) \,,
\end{align}
and the energy density of each component is given by $\rho_a=n^{\mu}n^{\nu}T^{(a)}_{\mu\nu}$ (cf.\ \eqref{eq:stress energy}).

\begin{figure}[]
\begin{centering}
\includegraphics[width=1\linewidth]{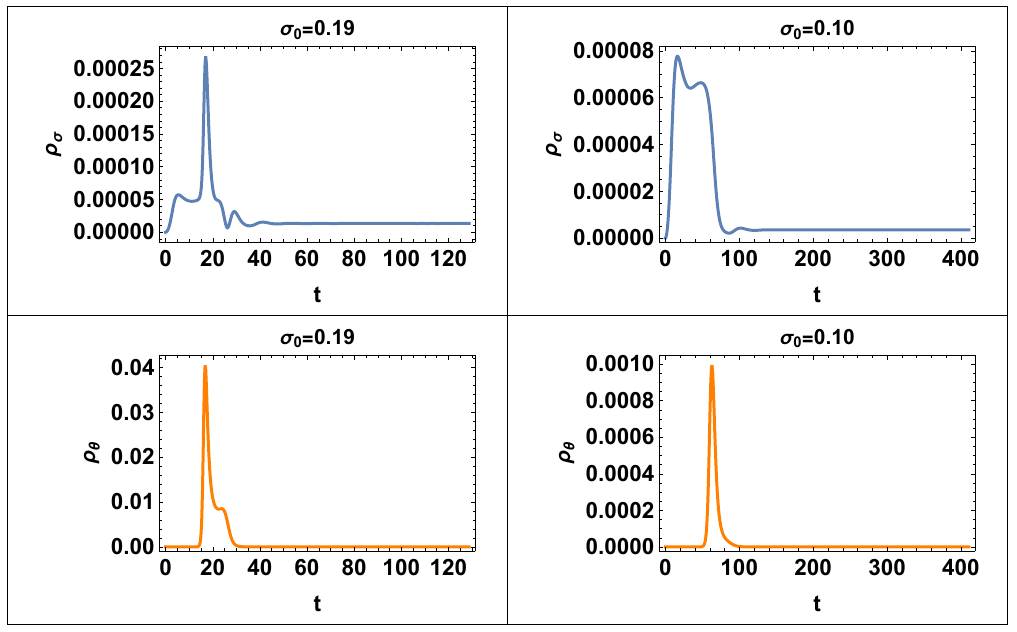}
\par\end{centering}
\caption{Time evolution of the energy densities $\rho_\sigma$ and $\rho_\theta$, evaluated at the event horizon, and using the initial values $\sigma_0=0.19$ and $\sigma_0=0.10$. One observes that both the radial and angular fields experience a peak in the energy density corresponding to the process of accretion onto the black hole.
}
\label{energydensity}
\end{figure}

\begin{figure}[]
\begin{centering}
\includegraphics[width=1\linewidth]{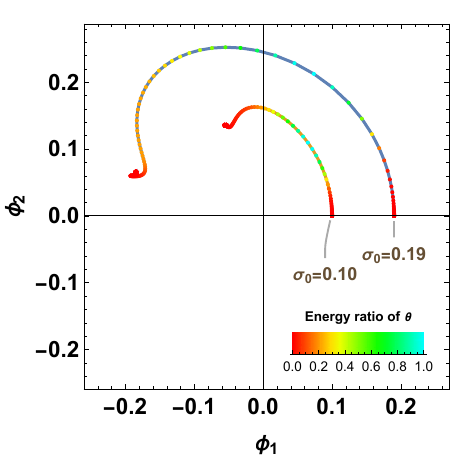}
\par\end{centering}
\caption{Trajectory of the horizon value of the scalar field, $\Phi_h$, in the $\phi_1-\phi_2$ plane, using the initial values $\sigma_0=0.19$ and $\sigma_0=0.10$. Dots are separated by equal time intervals, and their color indicates the ratio of $\rho_\theta$ to its maximum value.
}
\label{phi1phi2energy}
\end{figure}

The time dependence of $\rho_\sigma$ and $\rho_\theta$ is shown in Fig.\ \ref{energydensity} for the same simulations considered in the main text (cf.\ in particular Fig.\ \ref{Energy}). The results make it clear that the peak energy density of the angular field dominates over the peak value for the radial field, at least by an order of magnitude. Also noteworthy is the fact that the peak in $\rho_\theta$ has a relatively short duration, compared to the typical timescale for the accretion process. This is further illustrated in Fig.\ \ref{phi1phi2energy}, which shows that most of the field trajectory is traversed in the comparatively short time interval during which the angular field dominates the energy budget.

\bibliographystyle{unsrt}
\bibliography{ref}

\end{document}